\documentclass[aps, prx, twocolumn, amssymb, amsmath, superscriptaddress]{revtex4-1}

\usepackage{bm}
\usepackage{times}
\usepackage{graphicx}
\usepackage[colorlinks,citecolor=blue,linkcolor=blue]{hyperref}
\usepackage{array}
\usepackage{float}
\usepackage{colortbl}
\usepackage{multirow}
\usepackage{tabularx}

\begin{document}

\title{High phonon-limited mobility of BAs under pressure}
\author{Yanfeng Ge}
\affiliation{State Key Laboratory of Metastable Materials Science and Technology \& Key Laboratory for Microstructural Material Physics of Hebei Province,
School of Science, Yanshan University, Qinhuangdao, 066004, China}

\author{Wenhui Wan}
\affiliation{State Key Laboratory of Metastable Materials Science and Technology \& Key Laboratory for Microstructural Material Physics of Hebei Province,
School of Science, Yanshan University, Qinhuangdao, 066004, China}

\author{Yong Liu}\email{yongliu@ysu.edu.cn}
\affiliation{State Key Laboratory of Metastable Materials Science and Technology \& Key Laboratory for Microstructural Material Physics of Hebei Province,
School of Science, Yanshan University, Qinhuangdao, 066004, China}
\date{\today}

\begin{abstract}

Recent experiment reports that unusual high thermal conductivity has been observed in cubic boron arsenide crystal (BAs). In order to expand the scope of future multifunctional applications, we use first-principles calculations to investigate the phonon-limited electronic transport and modulation effect of pressure in BAs family. In the case of electron doping, BAs, AlAs and AlSb exhibit the coupling between electron and high frequency optical phonons, distinguished from the low frequency acoustical phonons in BSb. Thereinto BAs has the weakest electron-phonon coupling under the electron doping thus has a high N-type carrier mobility of 1740 cm$^{2}$/V$\cdot$s. After the introduction of pressure, phonon spectra has more obvious change than the electronic structure. The phonon hardening under the pressure gives rise to the weakening of electron-phonon coupling. It is obtained that the pressure of 50 GPa can improve the N-type carrier mobility of BAs up to 4300 cm$^{2}$/V$\cdot$s.

\end{abstract}

\maketitle

\section{Introduction}

In the past few decades, III-V semiconductors have generated intensive research interest owing to their special physical properties, and are widely expected to candidates for the next-generation devices of transistor~\cite{Ko2010,Alamo2011}, photovoltaic~\cite{Yoon2010,Wallentin2013} and light-emitting diode~\cite{Nakamura1994}. For example, the large electron-induced refractive-index change and low carrier-plasma absorption make III-V semiconductors have much larger transmission capacity than silicon photonics Mach-Zehnder modulators~\cite{Hiraki2017,Hu2018}. In addition, the indirect band gap of silicon, tremendous limitation of Si-based optoelectronic devices, can be overcome by the direct bandgap of some III-V semiconductors, which have also been in the development stage~\cite{Martensson2004,Chen2011,Ren2013,Wang2015,Kim2016,Mayer2016,Schuster2017}. Gallium arsenide (GaAs) is just outstanding among III-V semiconductors. It has a theoretical maximum one-sun energy conversion efficiency of 33.5\%~\cite{Green2014}, which can be further improved by the GaAs-based lattice-mismatched III-V semiconductors~\cite{Dimroth2014}. The researches about GaAs also refer to the spintronics~\cite{Ramsteiner2002,Dzhioev2002,Jungwirth2006,Jungwirth2014}, such as the spin Hall effect~\cite{Engel2005}, ferromagnetic order~\cite{Nazmul2005} and dynamic nuclear polarization~\cite{Nichol2015}. Furthermore, the Dirac bands have been observed in artificial graphene of GaAs quantum wells~\cite{Wang2018}, which spreads its area of application.

Very recently, cubic boron arsenide crystal BAs, one member of III-V semiconductors, has been experimentally isolated by the chemical vapor transport method with high purity source materials and has good prospects in research owing to its high thermal conductivity~\cite{Li2018,Kang2018,Tian2018}. The average bulk value of $\sim$1000 W/m$\cdot$K at room temperature is far above that of common metal materials and semiconductors in current main heat-removal systems, such as the value of 400 W/m$\cdot$K in copper. By comparing experiments with previous theoretical predictions~\cite{Lindsay2013,Broido2013,Lindsay2008,Feng2017}, the most compelling reason of high thermal conductivity is the large acoustic-optic gap in the phonon spectrum, which effectively limits the three-phonon scattering process. In short, high thermal conductivity makes BAs have immense potential in the next-generation electronic, optoelectronic and other devices.

The electronic transport, one of important physical properties, is worth serious study in order to realize the future application potential of BAs. Due to the high carrier mobility in other III-V semiconductors, the similar high-performance electronic transport is naturally expected in BAs. Here we investigate the influence of electron-phonon coupling on the electronic transport in BAs family (BAs, BSb, AlAs, and AlSb) as well as the modulation effect of pressure. The calculations of electronic structures demonstrate that all four materials are the semiconductors with indirect bandgap. After electron doping, the main electron-phonon coupling occurs at the high frequency optic phonon modes in BAs, AlAs, and AlSb, but the acoustic phonon modes of BSb. Among them, BAs has the weakest electron-phonon coupling. For the case of hole doping, the four materials all have strong electron-phonon coupling, and only focus on the high frequency optical phonon modes around $\Gamma$ point. By comparing all the various cases, the N-type carrier mobility of BAs is as high as 1740 cm$^{2}$/V$\cdot$s. In addition, the introduction of pressure mainly affect the phonon with frequency increasing, thus make electron-phonon weaken. It is very important that the mobility of N-type carrier in BAs is up to 4300 cm$^{2}$/V$\cdot$s under the pressure of 50 GPa.


\section{Methods}

The calculation is based on the semiclassical Boltzmann transport theory. Details of theory can be found in Ref.~\onlinecite{allen1978}.
The transport electron-phonon coupling constant $\lambda_{tr}$ can be obtained by,
\begin{eqnarray}
\lambda_{tr}=2\int^{\infty}_{0}\omega^{-1}\alpha^2_{tr}\rm{F}(\omega)\rm{d}\omega,
\label{eq:lambda}
\end{eqnarray}
where $\alpha^2_{tr}\rm{F}(\omega)$ is transport spectral function~\cite{allen1978},
\begin{eqnarray}
\alpha^2_{tr}\rm{F}(\alpha, \beta, \omega)=\alpha^2_{out}\rm{F}(\alpha, \beta, \omega)-\alpha^2_{in}\rm{F}(\alpha, \beta, \omega)
\label{eq:atr}
\end{eqnarray}

\begin{eqnarray}
	\begin{aligned}
		\alpha^2_{out(in)}\rm{F}(\alpha, \beta, \omega)=
		\frac{1}{N(E_F)\langle {v_{\alpha}(\epsilon_{F})} \rangle\langle {v_{\beta}(\epsilon_{F})} \rangle}\\
{\sum_{\substack{\mathbf q,\nu}}}{\sum_{\substack{i,j,\mathbf{k},\mathbf{k'}}}}
\delta(\omega-\omega_{\mathbf q,\nu})
		{\vert{M^{\nu}_{i\mathbf k, j\mathbf{k'}}}\vert}^2\\
		{v_{\alpha}(\mathbf{k})}{v_{\beta}(\mathbf{k^{(')}})}
		\delta(\epsilon_{i\mathbf k}-\epsilon_{F})
		\delta(\epsilon_{j\mathbf{k'}}-\epsilon_{F})
	\end{aligned}
	\label{eq:afw}
\end{eqnarray}
where $\alpha$ ($\beta$) indicating the directions and set to $x$ in the isotropous BAs family. The relaxation time $\tau$ can be derived by solving the Boltzmann equation in the lowest-order variational approximation (LOVA) as,
\begin{eqnarray}
\begin{aligned}
\tau^{-1}=
(\frac{\textstyle4\pi{k_B}T}{\hbar})
\int\frac{d\omega}{\omega}\frac{\tilde{\omega}^2}{\sinh^2\tilde{\omega}}\alpha^2_{tr}\rm{F}(\omega),
\end{aligned}
\label{eq:tau}
\end{eqnarray}
Thus, the temperature dependence of mobility ${\mu}(T)$ can be obtained by
\begin{eqnarray}
\mu(T)=\frac{2e {\emph N_ \emph F} \langle{v^2_x}\rangle }{n\textstyle{V_{cell}}} \tau,
\label{eq:mu}
\end{eqnarray}
where $\langle{v^2_x}\rangle$ is the average square of the Fermi velocity along the $x$ direction and $n$ is the carrier doping concentration.

\begin{figure}[htp!]
\centerline{\includegraphics[width=0.45\textwidth]{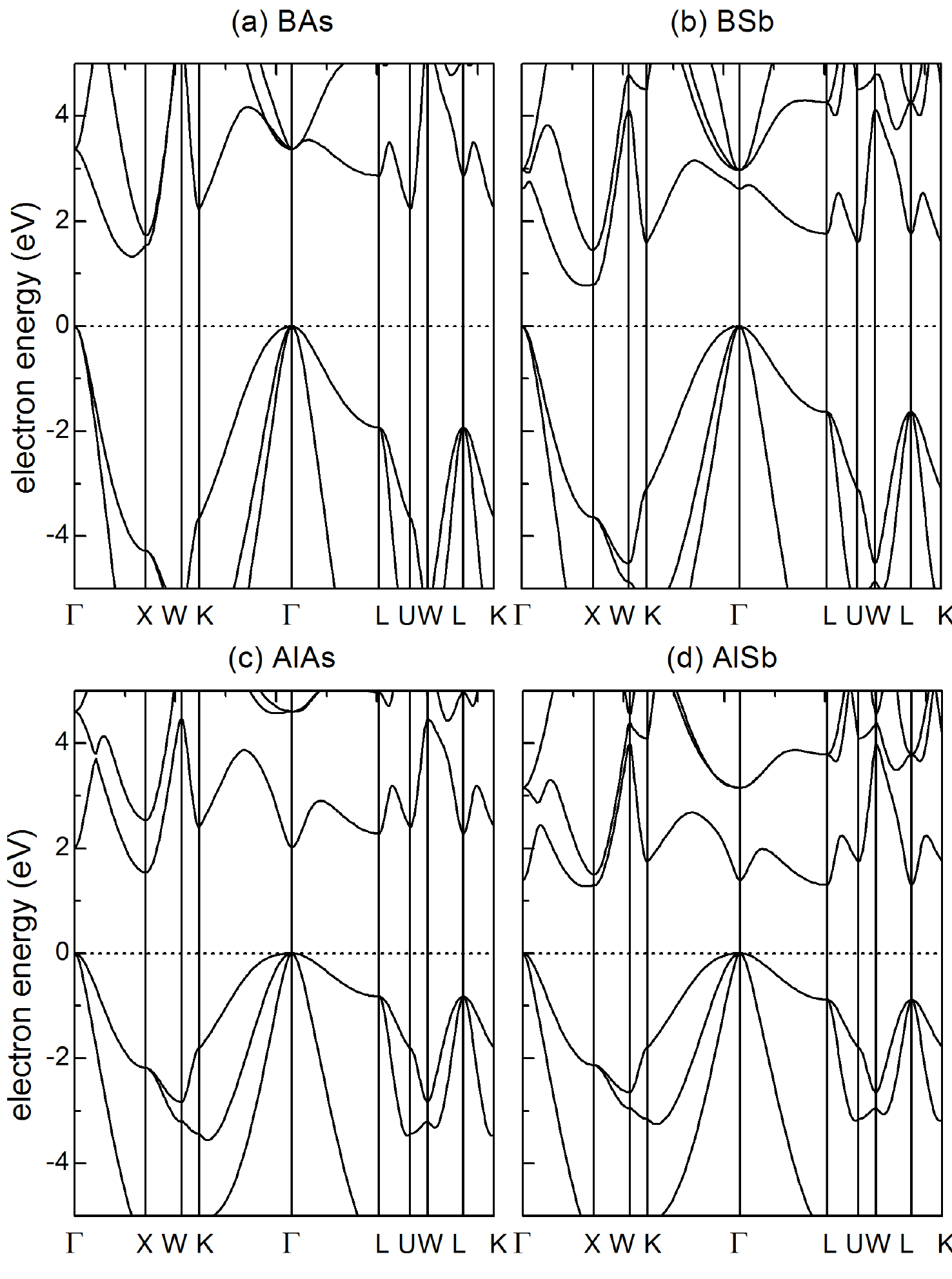}}
\caption{{\bf Band structures of BAs family} (a) Band structure of BAs, (b) BSb, (c) AlAs, and (d) AlSb with the indirect bangdap.
\label{fig:1-band}}
\end{figure}

Technical details of the calculations are as follows. All calculations, including the electronic structures, the phonon spectra, and the electron-phonon coupling, were carried out using the ABINIT package~\cite{Gonze19971,Gonze19972,Gonze2005,Gonze2009} with the local-density approximation (LDA). The ion and electron interactions are treated with the Hartwigsen-Goedecker-Hutter (HGH) pseudopotentials~\cite{Hartwigsen1998}.
The kinetic energy cutoff of $550$~eV and the Monkhorst-Pack $k$-mesh of 24$\times$24$\times$24 were used in all calculations about the electronic ground-state properties. The phonon spectra and the electron-phonon coupling were calculated on a 8$\times$8$\times$8 $q$-grid using the density functional perturbation theory (DFPT)~\cite{Baroni2001}.
Because of the original semiconductor of BAs, carrier doping was necessary for the study of electronic transport properties and tuned by shifting the Fermi level in the rigid-band approximation~\cite{Stern1967}.

\section{Results}

\subsection{Phonon-limited carrier mobility}

\begin{figure}[htp!]
\centerline{\includegraphics[width=0.45\textwidth]{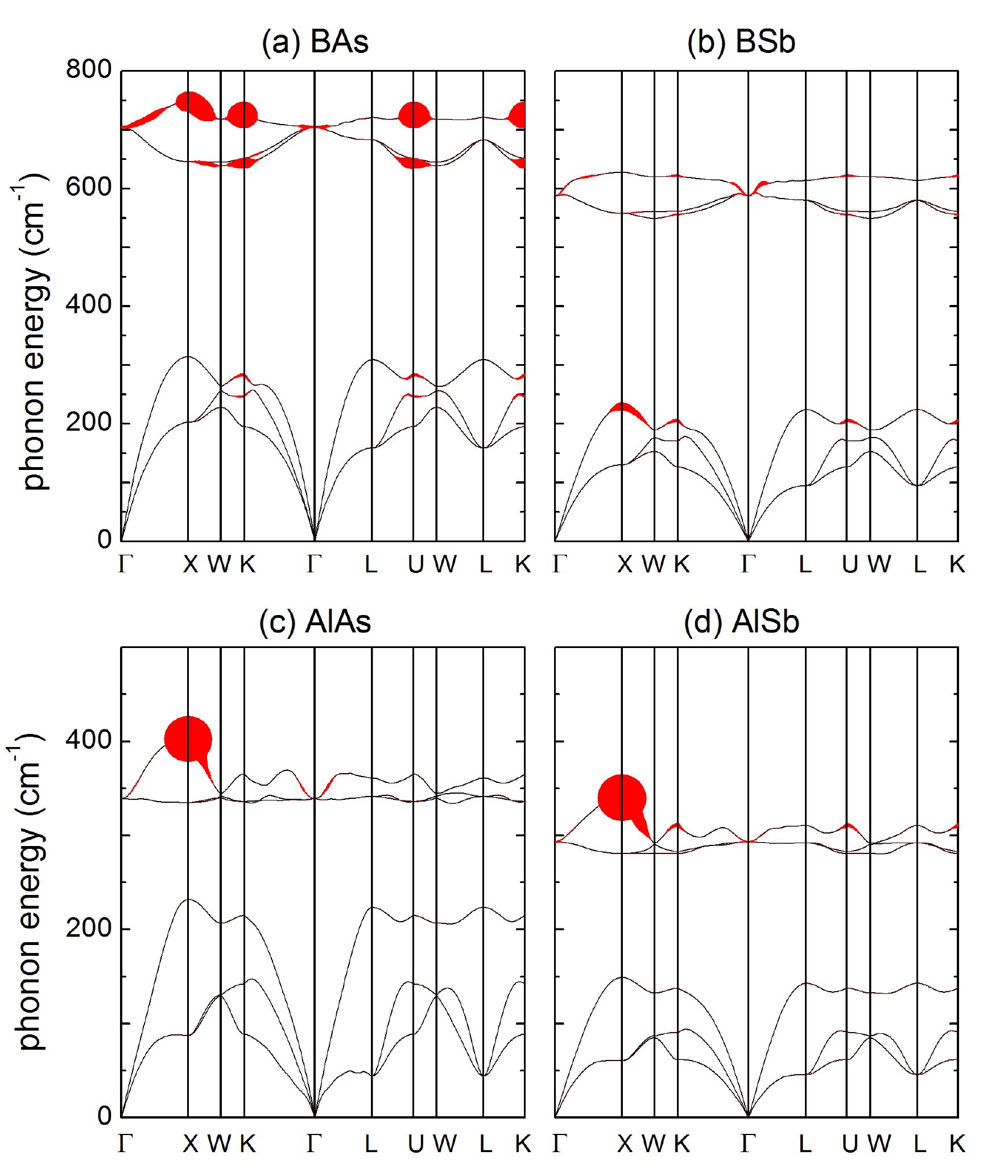}}
\caption{{\bf Phonon spectra of electron-doping BAs family.} (a) Phonon spectra of BAs, (b) BSb, (c) AlAs, and (d) AlSb with phonon linewidths, marked by the red error bar.
\label{fig:2-phonon-elec}}
\end{figure}

The lattice structure of BAs family is a face-centered cubic structure with space group of F$\bar{4}$3m as same as GaAs. In our calculations, the lattice constants are 4.81, 5.32, 5.70, and 6.28 \AA \ for BAs, BSb, AlAs, and AlSb, respectively. The band structures show that they all have indirect bandgap with valence band maximum (VBM) at $\Gamma$ point and the conduction band minimum (CBM) around X point [Fig.~\ref{fig:1-band}]. As the increase of atomic amass, the value of indirect bandgap change significantly and the conduction band around $\Gamma$ point also has a massive drop, as summarized in Tab.~\ref{tab:bandgap}. The main reason is that the elongate lattice constant for the cases of heavy atom makes energy of anti-bonding state at conduction band around $\Gamma$ point reduce, such as, AlSb has been closer to direct bandgap [Fig.~\ref{fig:1-band}d].

\begin{table}[htp!]
\caption{Bandgap of BAs family in unit of eV at zero pressure and 50 GPa. $\Delta$E$_{\rm gap}$ is the global indirect bandgap and $\Delta$E$_{\Gamma}$ is the local direct bandgap at $\Gamma$ point.}
\begin{tabular*}{6cm}{@{\extracolsep{\fill}}ccccccccc}
\hline\hline
& \multicolumn{2}{c}{$\Delta$E$_{\rm gap}$ } & \multicolumn{2}{c}{$\Delta$E$_{\Gamma}$ }\\
\hline
         system & 0 GPa & 50 GPa & 0 GPa & 50 GPa \\
\hline   BAs    & 1.32  & 1.27 & 3.37 & 3.46 \\
         BSb    & 0.78  & 0.70 & 2.62 & 3.11 \\
         AlAs   & 1.54  & 1.41 & 2.02 & 2.62 \\
         AlSb   & 1.27  & 1.16 & 1.39 & 2.00 \\
\hline \hline
\end{tabular*}
\label{tab:bandgap}
\end{table}

\begin{figure}[htp!]
\centerline{\includegraphics[width=0.45\textwidth]{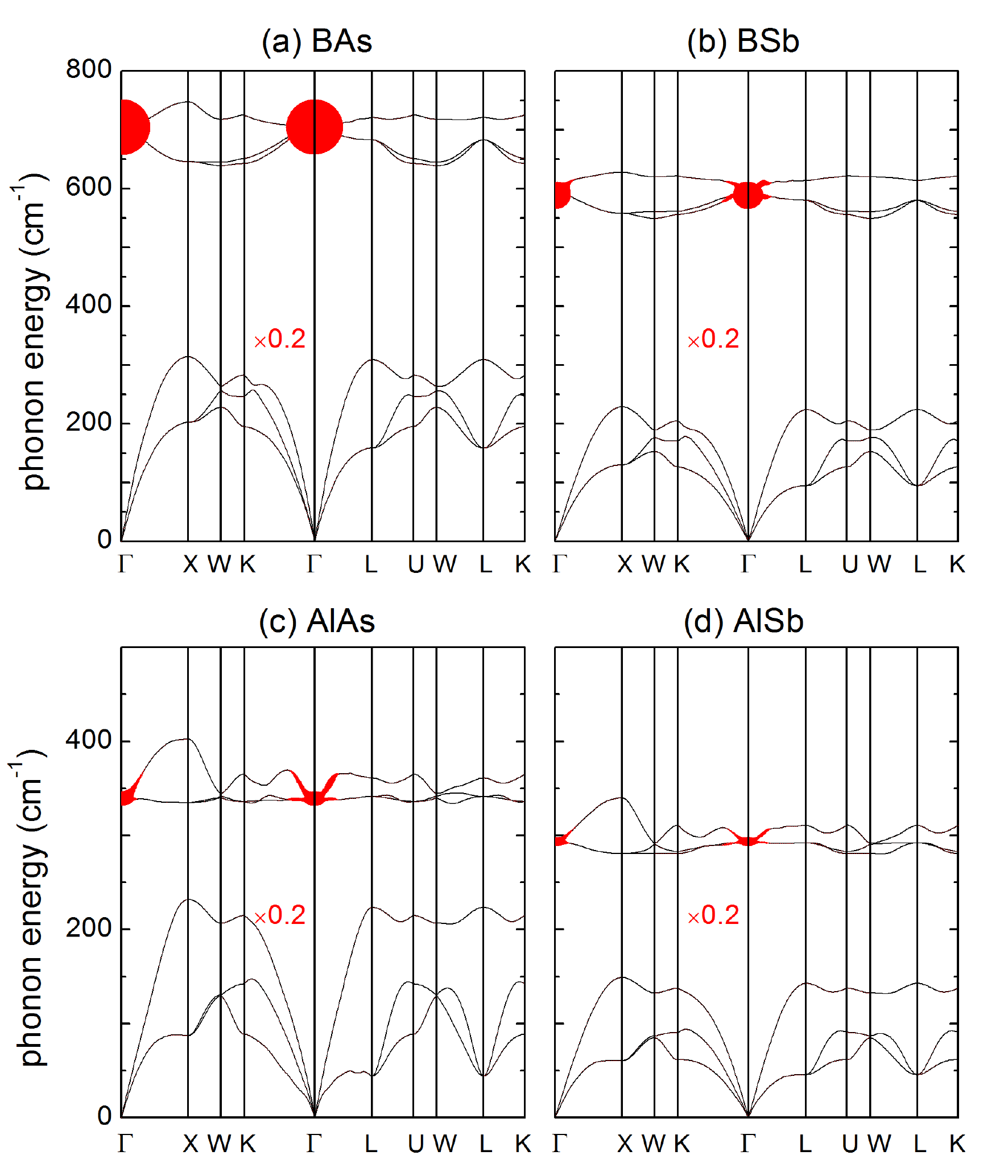}}
\caption{{\bf Phonon spectra of hole-doping BAs family.} (a) Phonon spectra of BAs, (b) BSb, (c) AlAs, and (d) AlSb with phonon linewidths, marked by the red error bar. And the magnitude of phonon linewidths are too large and thus plotted with one fifth of the real values.
\label{fig:3-phonon-hole}}
\end{figure}

In order to study the electronic transport in semiconductor BAs family, N-type and P-type carriers are both considered in the present work. Figures.~\ref{fig:2-phonon-elec} and ~\ref{fig:3-phonon-hole} are the phonon spectra with phonon linewidths for electron and hole doping, respectively. First, the absent of imaginary frequency in four materials ensures the dynamics stability of lattice structures. And it is natural that there are also significant effect of atomic amass on the lattice vibration. When the atomic amass increases, the phonon frequencies decrease remarkably, such as the highest frequency dropping from 760 cm$^{-1}$ (BAs) to 350 cm$^{-1}$ (AlSb) [Fig.~\ref{fig:2-phonon-elec}]. In addition, the size of red error bar plots the magnitude of the phonon linewidths, which denotes the contributors of electron-phonon coupling. For the electron-doping case, it is evident that three high frequency optical phonon modes of BAs have much larger phonon linewidths than that of acoustic phonon modes.
Because the CBM of BAs deviates X point slightly, the phonon-assisted electronic scattering processes are various and complex, thus the large phonon linewidths locate around X, K and U points [Fig.~\ref{fig:2-phonon-elec}(a)]. But the exact opposite occurs in BSb, where the acoustic phonon modes around 200 cm$^{-1}$ has large phonon linewidths, as shown in Fig.~\ref{fig:2-phonon-elec}(b). Moreover, AlAs and AlSb both have the large phonon linewidths locating at the optical phonon modes only around X point. The results of hole-doping case reveal the larger phonon linewidths than that of electron doping [Fig.~\ref{fig:3-phonon-hole}]. The region of dominating electron-phonon coupling also has a very significant change and focus on the $\Gamma$ point in the all four cases.

\begin{figure}[htp!]
\centerline{\includegraphics[width=0.45\textwidth]{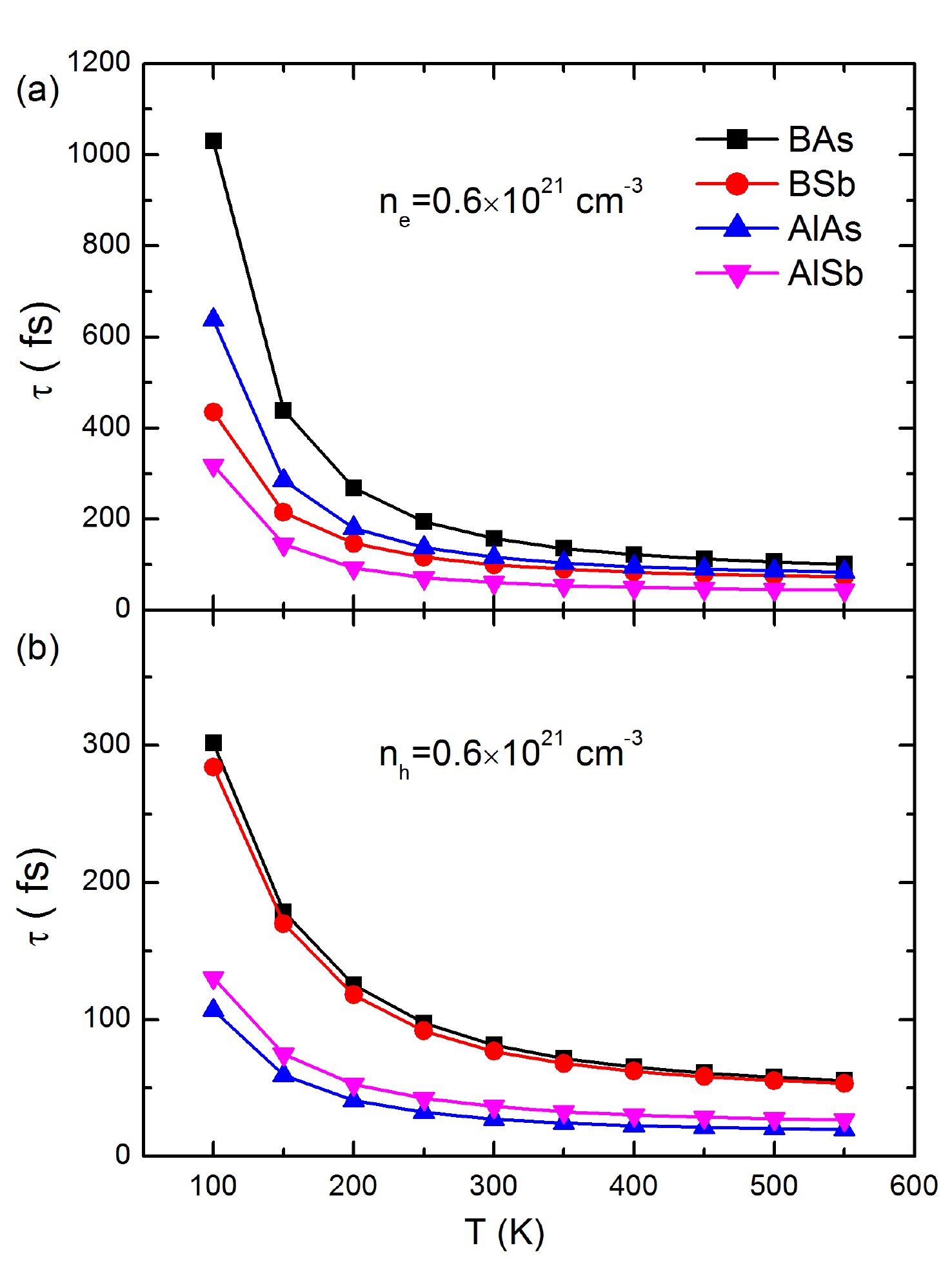}}
\caption{{\bf relaxation time of BAs family.} (a) Relaxation time $\tau$ as a function of the temperature at the doping concentration n = 0.6$\times$10$^{21}$ cm$^{-3}$ of N-type carriers and (b) N-type carriers.
\label{fig:4-tau}}
\end{figure}

Basing on the above electron-phonon coupling, we obtain the relaxation time $\tau$ in the Boltzmann theory of electronic transport. Figure.~\ref{fig:4-tau}(a) shows the relaxation time $\tau$ of N-type carriers versus temperature at n$_e$ = 0.6$\times$10$^{21}$ cm$^{-3}$. It is observed that BAs has the longest carrier relaxation time and AlSb has the shortest value in the temperature range of 100 $\sim$ 550 K. At room temperature, the N-type carriers in BAs has relaxation time $\tau$ of $\sim$ 200 fs even up to the order of picoseconds at 100 K. The main reason is analyzed by using phonon spectra with phonon linewidths. For the BAs, AlAs, and AlSb, although the optical phonon modes contribute the main electron-phonon coupling, the lower frequency from the heavier atomic amass (AlAs and AlSb) lead to the increase of integral transport spectral function in $\lambda_{tr}$ and the inverse of relaxation time $\tau^{-1}$ [Eq.~\ref{eq:lambda} and~\ref{eq:tau}]. In a similar way, the relatively small phonon linewidths of the acoustic phonon modes in the low frequency region make BSb also have short relaxation time. In addition, the relaxation times of P-type carriers are much shorter than that of N-type carriers [Fig.~\ref{fig:4-tau}] because of the strong coupling between the electrons and the phonons around $\Gamma$ points [Fig.~\ref{fig:3-phonon-hole}]. And the void between variation curve of $\tau$ for B-systems and Al-systems under the hole-doping condition also derives the shift downward in frequency due to the heavy atom.

\begin{figure}[htp!]
\centerline{\includegraphics[width=0.45\textwidth]{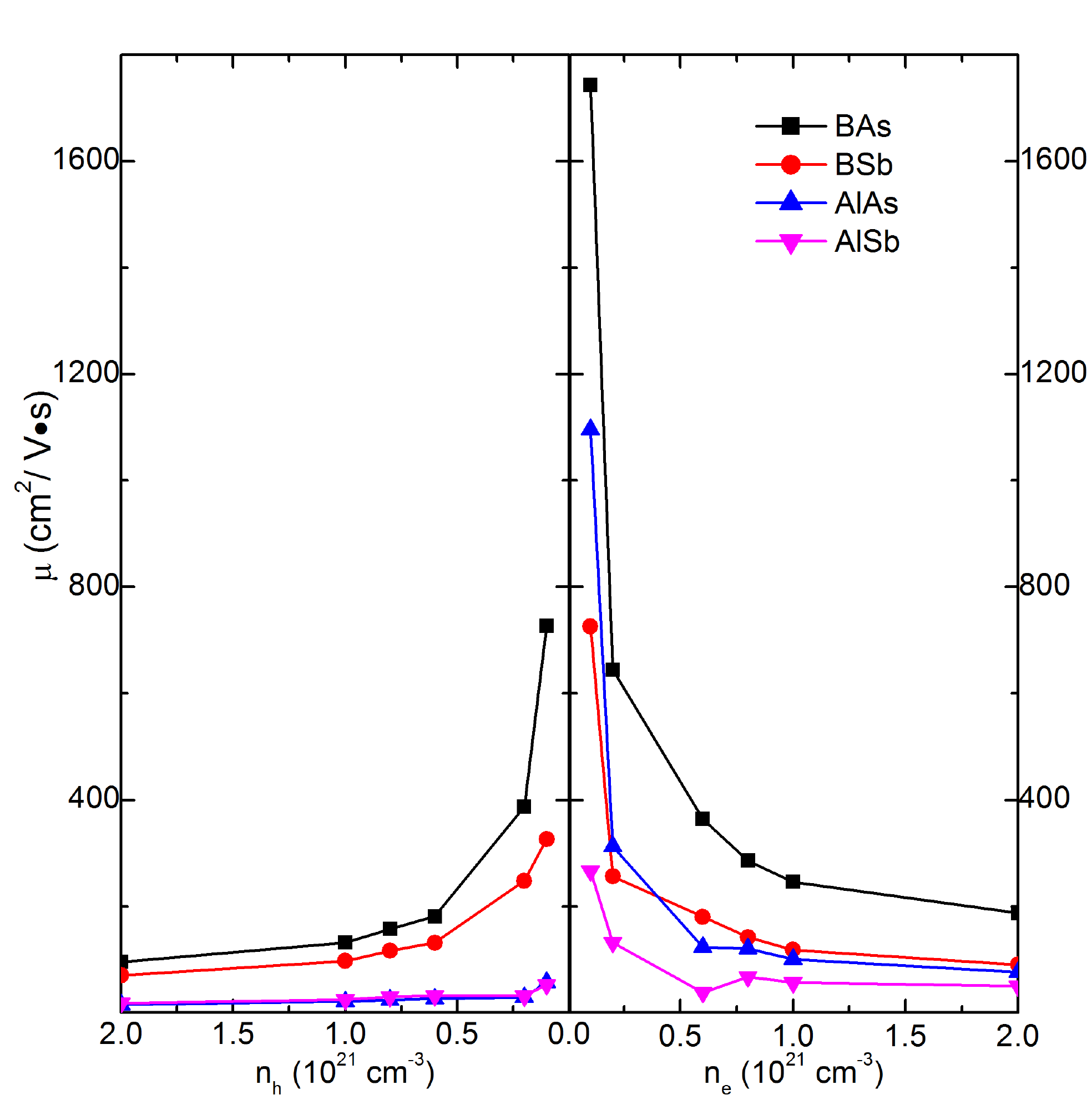}}
\caption{{\bf Mobility of BAs family.} Mobility $\mu$ as a function of the doping concentration of N-type carriers and N-type carriers.
\label{fig:5-mu}}
\end{figure}

\begin{figure}[htp!]
\centerline{\includegraphics[width=0.45\textwidth]{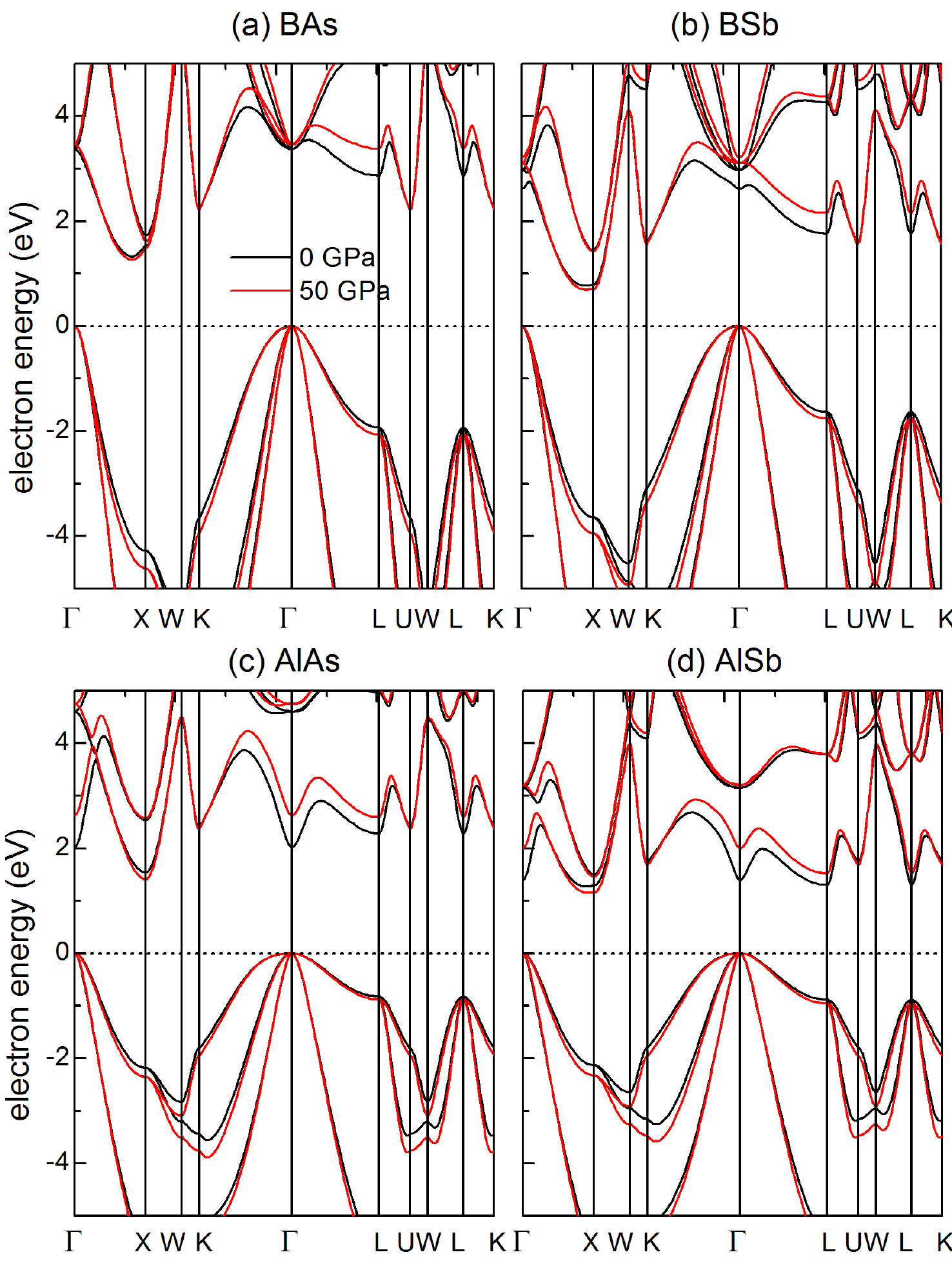}}
\caption{{\bf Band structures of BAs family under pressure} (a) Band structure of BAs, (b) BSb, (c) AlAs, and (d) AlSb. The band structures under pressure of 50 GPa (zero pressure) are plotted by the red (black) line.
\label{fig:6-band-gpa}}
\end{figure}

According to the equation of mobility [Eq.~\ref{eq:mu}], we estimate the electronic transport properties of electron and hole doping in BAs family at room temperature. As shown in Fig.~\ref{fig:5-mu}, the mobility $\mu$ of N-type carriers has much higher values than that of P-type carriers, due to the weaker electron-phonon coupling of electron doping [Fig.~\ref{fig:2-phonon-elec} and~\ref{fig:3-phonon-hole}]. And $\mu$ has a monotonous downward trend versus the carrier concentration, as the result of the more density of electronic states (DOS) more stronger electron-phonon coupling at the high doping concentration. It is found that N-type and P-type carriers of BAs have $\mu$ of 1740 and 726 cm$^2$/V$\cdot$s at n = 0.1$\times$10$^{21}$ cm$^{-3}$, respectively. The results of electron-doping BAs also in agreement with the previous work~\cite{Liu2018}, but a difference exists in the case of hole doping, which is higher performance than electron doping in Ref.~\onlinecite{Liu2018}. We hold the opinion that the doping concentration is key factor. Considering the high degenerate at VBM [Fig.~\ref{fig:1-band}], the high doping concentration will lead to the large DOS so increases the electron-phonon coupling obviously. For other cases in the present work, the stronger electron-phonon coupling than BAs lead to the low values of mobility, such as the 260 cm$^2$/V$\cdot$s of N-type carrier in AlSb when n$_e$ = 0.1$\times$10$^{21}$ cm$^{-3}$.

\begin{figure}[htp!]
\centerline{\includegraphics[width=0.45\textwidth]{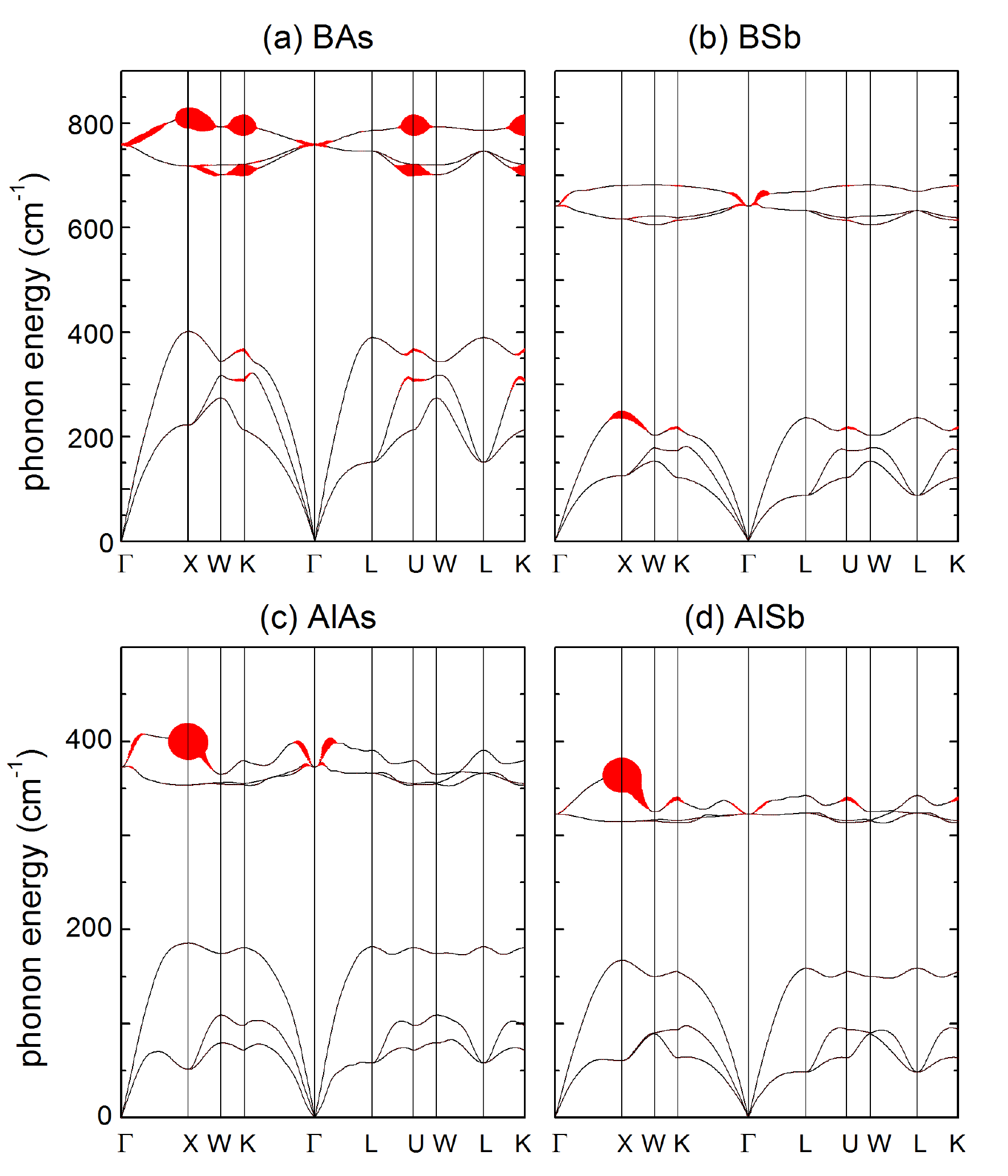}}
\caption{{\bf Phonon spectra of electron-doping BAs family under pressure.} (a) Phonon spectra of BAs, (b) BSb, (c) AlAs, and (d) AlSb at pressure of 50GPa, with phonon linewidths, marked by the red error bar.
\label{fig:7-phonon-elec-gpa}}
\end{figure}

\begin{figure}[t!]
\centerline{\includegraphics[width=0.45\textwidth]{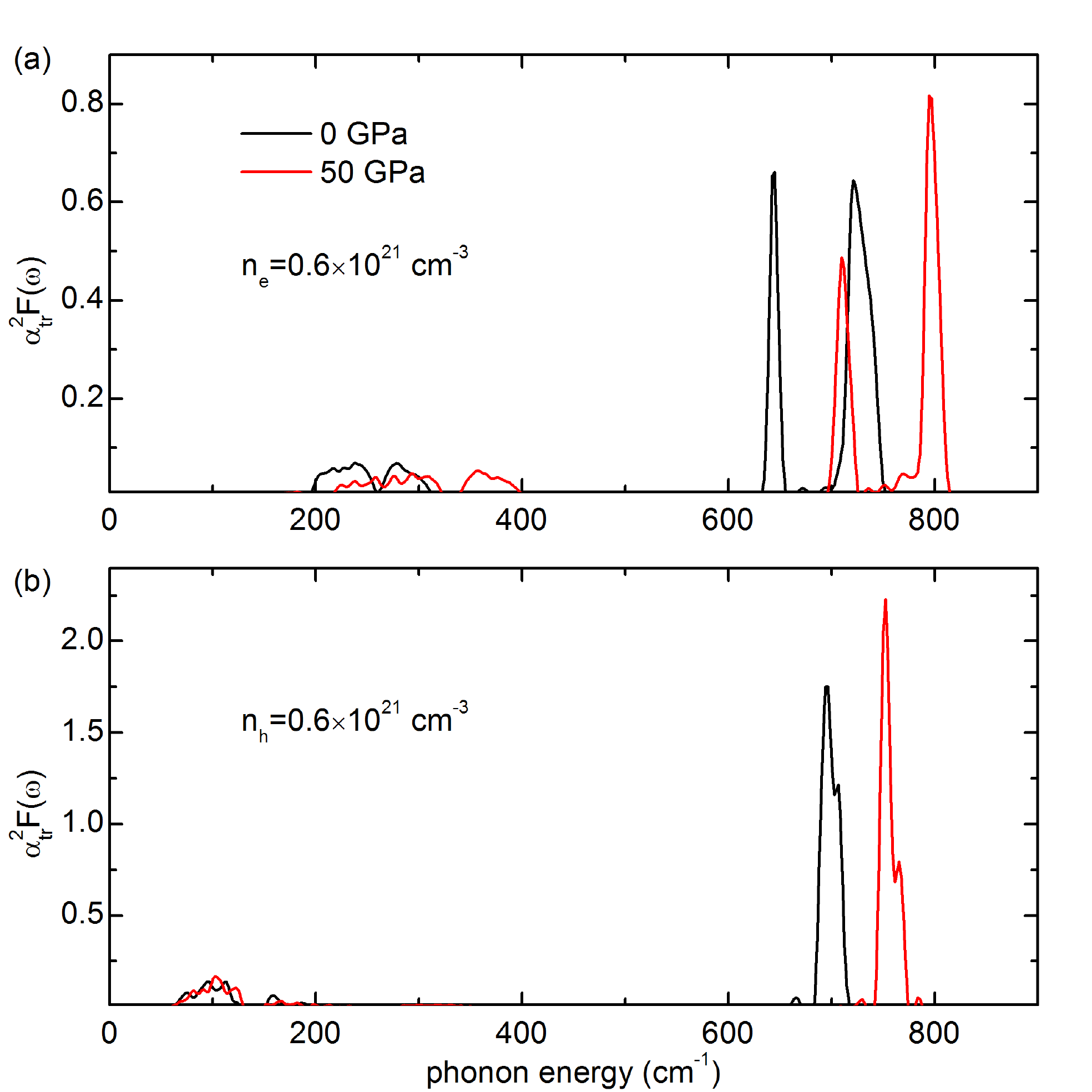}}
\caption{{\bf Transport spectral function of BAs under pressure.} (a) Transport spectral function $\alpha^2_{tr}\rm{F}(\omega)$ of BAs with electron-doping and (b) hole doping.
\label{fig:8-afw}}
\end{figure}

\subsection{Pressure effect}

In order to improve the electronic transport of BAs family, we also systematically study the effect of pressure on the electron-phonon coupling. In the band structures, it's worth noting that the effect of pressure is particularly reflected in the conduction band around $\Gamma$ point because the change of lattice constant under pressure influences the anti-bonding state [Fig.~\ref{fig:6-band-gpa}]. But the global indirect bandgap is impacted minimally [Tab.~\ref{tab:bandgap}].

\begin{figure}[t!]
\centerline{\includegraphics[width=0.5\textwidth]{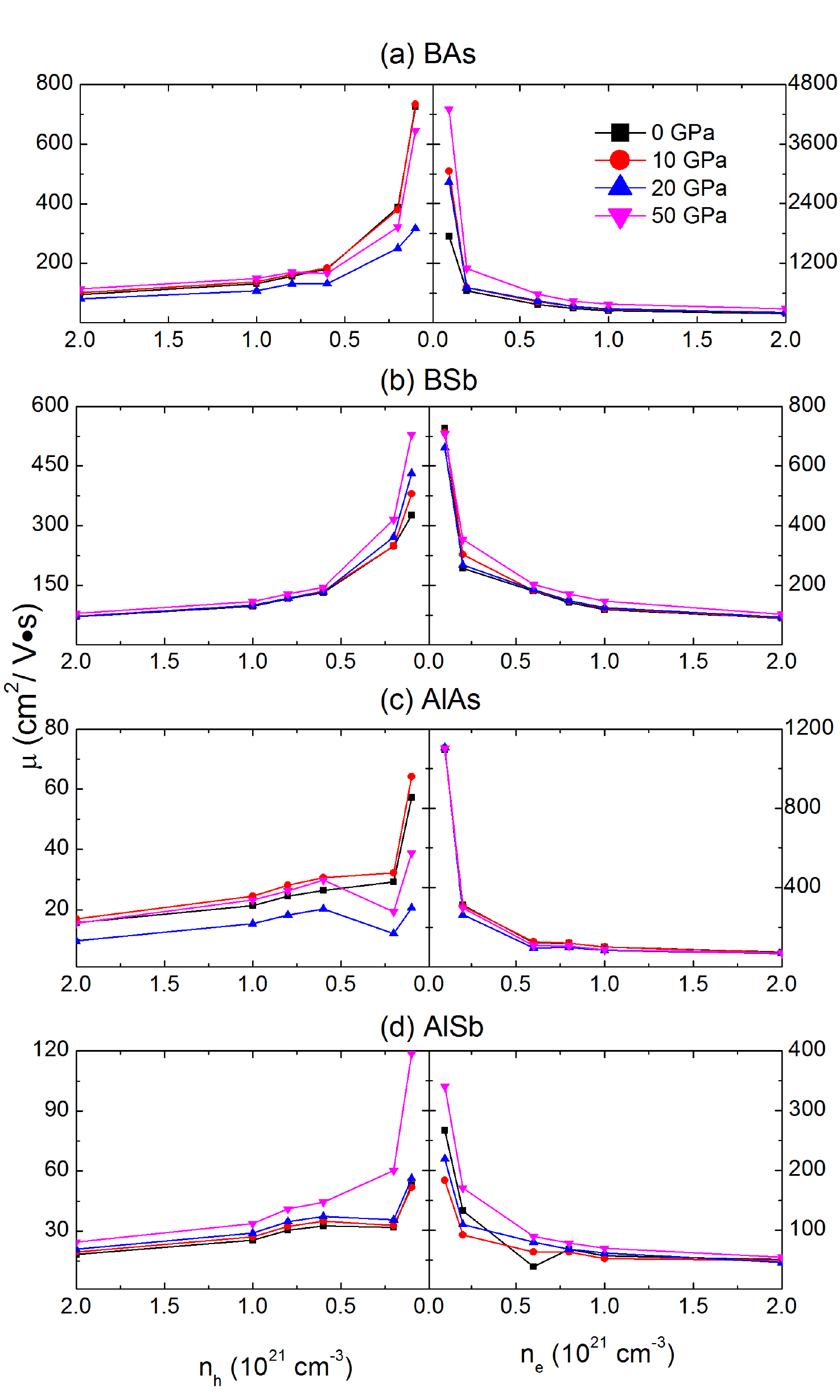}}
\caption{{\bf Mobility of BAs family under pressure.} Mobility $\mu$ as a function of the doping concentration of N-type carriers and N-type carriers.
\label{fig:9-mu-gpa}}
\end{figure}

However, the pressure has greater influence on the phonon spectra. First, the bond length shortens under pressure, also covalent bond energy increasing, which makes frequencies rise in the entire phonon spectra, as shown in Fig.~\ref{fig:7-phonon-elec-gpa}. Due to the insignificant change of band structure, the regions of large phonon linewidths of electron-doping BAs under 50 GPa also focus on the three points of X, K, and U in the high frequency optical phonon modes. For other materials, the region of electron-phonon coupling is same as the case of zero pressure [Fig.~\ref{fig:7-phonon-elec-gpa}], which also occurs in the case of hole doping. Contrary to the effect of heavy atom, the increase of phonon frequency under pressure is to the benefit of weakening the electron-phonon coupling [Eq.~\ref{eq:lambda}], so the pressure further boosts the electronic transport performance.

As shown in Fig.~\ref{fig:8-afw}, the blue shift of peaks of high frequency in the transport spectral function is in consistent with the phonon hardening under pressure. And for the electron doping, the peak at low frequency also shift clearly, just opposite to the little change in the hole doping. So the enhance-effect of pressure on the N-type carrier mobility is more noticeable than that of P-type carrier [Fig.~\ref{fig:9-mu-gpa}]. As plotted in Fig.~\ref{fig:9-mu-gpa}(a), the pressure of 50 GPa can improve the N-type carrier mobility of BAs from 1740 up to 4300 cm$^2$/(V$\cdot$s) when n$_e$ = 0.1$\times$10$^{21}$ cm$^{-3}$. There are also promotions of mobility for other cases by the pressure, which are much more inconspicuous than that of BAs yet.


\section{Discussion}

In the present work, we have investigated the phonon-limited electronic transport in BAs family (BAs, BSb, AlAs, and AlSb). The electronic structures shows that all four materials have indirect bandgap in the visible region thus electron and hole doping are needed to be considered in order to obtain the carrier mobility. First, in the condition of electron doping, it is found that the main electron-phonon coupling occurs at the high frequency optic phonon modes in BAs, AlAs, and AlSb, but the acoustic phonon modes of BSb. Because the heavy atom make the phonon frequency drop (AlAs and AlSb) or low frequency phonons contribute the electron-phonon coupling (BSb), the three materials apart from BAs have strong electron-phonon coupling, which also occurs the four materials under the hole doping. The results shows that BAs has the N-type carrier mobility of 1740 cm$^{2}$/V$\cdot$s under pressureless condition.

In addition, the introduction of pressure has little overall effect on the electronic structure around the bandgap. But the enhancement of bonding energy under pressure leads to the phonon hardening, which is favorable to weaken electron-phonon coupling. Thus, it's discovered that the pressure can boost the electronic transport performance of four materials even further. Especially the N-type carrier mobility of BAs is as high as 4300 cm$^{2}$/V$\cdot$s under the pressure of 50 GPa, far exceeding the vast majority found semiconductors, which is great significance to the current semiconductor industry. Furthermore, the phonon spectra under pressure also has a large frequency gap between acoustic and optic phonons, which is the key factor for the high thermal conductivity in BAs as previously mentioned, so that pressure does not negatively impact the thermal transport. The present results illustrate that the effective enhancement of pressure on mobility in BAs family still with high thermal conductivity and provide basis of theoretical calculation for the future experiments.

\begin{acknowledgments}
This work was supported by the NSFC (Grants No.11747054), the Specialized Research Fund for the Doctoral Program of Higher Education of China (Grant No.2018M631760), the Project of Heibei Educational Department, China (No. ZD2018015 and QN2018012), and the Advanced Postdoctoral Programs of Hebei Province (No.B2017003004).
\end{acknowledgments}


\begin{thebibliography}{100}
\expandafter\ifx\csname url\endcsname\relax
  \def\url#1{\texttt{#1}}\fi
\expandafter\ifx\csname urlprefix\endcsname\relax\def\urlprefix{URL }\fi
\providecommand{\bibinfo}[2]{#2}
\bibitem{Ko2010}H. Ko, K. Takei, R. Kapadia, S. Chuang, H. Fang, P. W. Leu, et al. Nature \textbf{468}, 286 (2010).
\bibitem{Alamo2011}J. A. Del Alamo Nature \textbf{479}, 317 (2011).
\bibitem{Yoon2010}J. Yoon, S. Jo, I. S. Chun, I. Jung, H. S. Kim, M. Meitl, et al. Nature \textbf{465}, 329 (2010).
\bibitem{Wallentin2013}J. Wallentin, N. Anttu, D. Asoli, M. Huffman, I. Aberg, et al. Science \textbf{339}, 1057 (2013).
\bibitem{Nakamura1994}S. Nakamura, T. Mukai, M. Senoh, Appl. Phys. Lett. \textbf{64}, 1687 (1994).
\bibitem{Hiraki2017}T. Hiraki, T. Aihara, K. Hasebe, K. Takeda, T. Fujii, et al. Nature Photonics \textbf{11}, 482 (2017).
\bibitem{Hu2018}H. Hu, F. Da Ros, M. Pu, F. Ye, K. Ingerslev, E. P. da Silva, et al. Nature Photonics \textbf{12}, 469 (2018).
\bibitem{Martensson2004}T. Martensson, C. P. T. Svensson, B. A. Wacaser, M. W. Larsson, et al. Nano Lett. \textbf{4}, 1987 (2004).
\bibitem{Chen2011}R. Chen, T. T. D. Tran, K. W. Ng, W. S. Ko, L. C. Chuang, et al. Nature Photonics \textbf{5}, 170 (2011).
\bibitem{Ren2013}F. Ren, K. Wei Ng, K. Li, H. Sun, et al. Appl. Phys. Lett. \textbf{102}, 012115 (2013).
\bibitem{Wang2015}Z. Wang, B. Tian, M. Pantouvaki, W. Guo, P. Absil, et al. Nature Photonics \textbf{9}, 837 (2015).
\bibitem{Kim2016}H. Kim, A. C. Farrell, P. Senanayake, W. J. Lee, et al. Nano Lett. \textbf{16}, 1833 (2016).
\bibitem{Mayer2016}B. Mayer, L. Janker, B. Loitsch, J. Treu, T. Kostenbader, et al. Nano Lett. \textbf{16}, 152 (2016).
\bibitem{Schuster2017}F. Schuster, J. Kapraun, G. N. Malheiros-Silveira, S. Deshpande, and C. J. Chang-Hasnain, Nano Lett. \textbf{17}, 2697 (2017).
\bibitem{Green2014}M. A. Green, K. Emery, Y. Hishikawa, W. Warta and E. D. Dunlop, Prog. Photovoltaics, \textbf{22}, 1 (2014).
\bibitem{Dimroth2014}F. Dimroth, M. Grave, P. Beutel, U. Fiedeler, C. Karcher, T. N. D. Tibbits, et al. Prog. Photovolt: Res. Appl. \textbf{22}, 277 (2014).
\bibitem{Ramsteiner2002}M. Ramsteiner, H. Y. Hao, A. Kawaharazuka, H. J. Zhu, M. Kastner, R. Hey, L. Daweritz, H. T. Grahn, and K. H. Ploog Phys. Rev. B \textbf{66}, 081304 (2002).
\bibitem{Dzhioev2002}R. I. Dzhioev, K. V. Kavokin, V. L. Korenev, M. V. Lazarev, B. Ya. Meltser, M. N. Stepanova, B. P. Zakharchenya, D. Gammon, and D. S. Katzer, Phys. Rev. B \textbf{66}, 245204 (2002).
\bibitem{Jungwirth2006}T. Jungwirth, J. Sinova, J. Masek, J. Kucera, and A. H. MacDonald, Rev. Mod. Phys. \textbf{78}, 809 (2006).
\bibitem{Jungwirth2014}T. Jungwirth, J. Wunderlich, V. Novak, K. Olejnik, B. L. Gallagher, R. P. Campion, K. W. Edmonds, A. W. Rushforth, A. J. Ferguson, and P. Nemec Rev. Mod. Phys. \textbf{86}, 855 (2014).
\bibitem{Engel2005}H.-A. Engel, B. I. Halperin, and E. I. Rashba, Phys. Rev. Lett. \textbf{95}, 166605 (2005).
\bibitem{Nazmul2005}A. M. Nazmul, T. Amemiya, Y. Shuto, S. Sugahara, and M. Tanaka, Phys. Rev. Lett. \textbf{95}, 017201 (2005).
\bibitem{Nichol2015}J. M. Nichol, S. P. Harvey, M. D. Shulman, A. Pal, V. Umansky, E. I. Rashba, B. I. Halperin, and A. Yacoby, Nat. Commun. \textbf{6}, 7682 (2015).
\bibitem{Wang2018}S. Wang, D. Scarabelli, L. Du, Y. Y. Kuznetsova, L. N. Pfeiffer, et al. Nature Nanotechnology \textbf{13}, 29 (2018).
\bibitem{Li2018}S. Li, Q. Zheng, Y. Lv, X. Liu, X. Wang, P. Y. Huang, D. G. Cahill, B. Lv, Science \textbf{10.1126/science.aat8982} (2018).
\bibitem{Kang2018}J. S. Kang, M. Li, H. Wu, H. Nguyen, Y. Hu, Science \textbf{10.1126/science.aat5522} (2018).
\bibitem{Tian2018}F. Tian, B. Song, X. Chen, N. K. Ravichandran, Y. Lv, et al., Science \textbf{10.1126/science.aat7932} (2018).
\bibitem{Lindsay2013}L. Lindsay, D. A. Broido, T. L. Reinecke, Phys. Rev. Lett. \textbf{111}, 025901 (2013).
\bibitem{Broido2013}D. A. Broido, L. Lindsay, and T. L. Reinecke, Phys. Rev. B \textbf{88}, 214303 (2013).
\bibitem{Lindsay2008}L. Lindsay, D. A. Broido, J. Phys. Condens. Matter \textbf{20}, 165209 (2008).
\bibitem{Feng2017}T. Feng, L. Lindsay, X. Ruan, Phys. Rev. B \textbf{96}, 161201 (2017).
\bibitem{Liu2018}T.-H. Liu, B. Song, L. Meroueh, Z. Ding, Q. Song, J. Zhou, M. Li, and G. Chen, Phys. Rev. B \textbf{98}, 081203(R) (2018).
\bibitem{allen1978} P. B. Allen, {Phys. Rev. B} \textbf{17}, 3725 (1978).
\bibitem{Gonze19971}X. Gonze, {Phys. Rev. B} \textbf{55}, 10337 (1997).
\bibitem{Gonze19972}X. Gonze, and C. Lee, {Phys. Rev. B} \textbf{55}, 10355 (1997).
\bibitem{Gonze2005}X. Gonze, G.-M. Rignanese, M. Verstraete, J.-M. Beuken, Y. Pouillon, et al. {Z. Kristallogr.} \textbf{220}, 558 (2005).
\bibitem{Gonze2009}X. Gonze, B. Amadon, P.-M. Anglade, J.-M. Beuken, F. Bottin, P. Boulanger, F. Bruneval, et al. {Comput. Phys. Commun.} \textbf{180}, 2582 (2009).
\bibitem{Hartwigsen1998}C. Hartwigsen, S. Goedecker, and J. Hutter, {Phys. Rev. B} \textbf{58}, 3641 (1998).
\bibitem{Allen1983}P. B. Allen, and B. Mitrovi$\acute{\rm c}$, {Solid State Physics} \textbf{37}, 1 (1983).
\bibitem{Allen1975}P. B. Allen, and R. C. Dynes, {Phys. Rev. B} \textbf{12}, 905 (1975).
\bibitem{Durajski}A. P. Durajski, R. Szcze$\acute{\rm s}$niak, and Y. Li, {Physica C} \textbf{515}, 1 (2015).
\bibitem{McMillan1968}W. L. McMillan, {Phys. Rev.} \textbf{167}, 331 (1968).
\bibitem{Allen1972}P. B. Allen, {Phys. Rev. B} \textbf{6}, 2577 (1972).
\bibitem{Allen1974}P. B. Allen, and R. Silberglitt, {Phys. Rev. B} \textbf{9}, 4733 (1974).
\bibitem{Grimvall}G. Grimvall, (North-Holland, Amster-dam, 1981).
\bibitem{Baroni2001} S. Baroni, S. D. Gironcoli, A. D. Corso and P. Giannozzi, {Rev. Mod. Phys.} \textbf{73}, 515 (2001).
\bibitem{Stern1967}E. A. Stern, Phys. Rev. 157, 544 (1967) .

\end{thebibliography}
\end{document}